\documentclass[preprint,5p,twocolumn]{elsarticle}

\usepackage{hyperref}
\usepackage{graphicx}
\usepackage{multirow}
\usepackage{soulutf8}
\usepackage{color}
\usepackage{subcaption}
\usepackage[normalem]{ulem}
\usepackage[dvipsnames]{xcolor}

\newcommand{\marev}[1]{\textcolor{black}{#1}}
\newcommand{\mirev}[1]{\textcolor{black}{#1}}

\graphicspath{{figures/}}
\journal{Computer Networks}
\biboptions{sort&compress}

\begin{document}

\begin{frontmatter}


\title{S-HIDRA: A blockchain and SDN domain-based architecture to orchestrate fog computing environments}


\author[add1]{Carlos Núñez-Gómez\corref{cauthor}}
\address[add1]{High-Performance Networks and Architectures Group (RAAP), Albacete Research Institute of Informatics (I3A), University of Castilla-La Mancha, 02071 Albacete, Spain}
\ead{carlos.nunez@uclm.es}
\cortext[cauthor]{Corresponding author}

\author[add2]{Carmen Carrión}
\ead{carmen.carrion@uclm.es}
\author[add2]{Blanca Caminero}
\ead{mariablanca.caminero@uclm.es}
\author[add2]{Francisco M. Delicado}
\ead{francisco.delicado@uclm.es}
\address[add2]{Department of Computing Systems, University of Castilla-La Mancha, 02071 Albacete, Spain}

\begin{abstract}
Fog computing arises as a complement to cloud computing where computing and storage are provided in a decentralized way rather than the centralized approach of the cloud paradigm. In addition, blockchain provides a decentralized and immutable ledger which can provide support for running arbitrary logic thanks to smart contracts. These facts can lead to harness smart contracts on blockchain as the basis for a decentralized, autonomous, and resilient orchestrator for the resources in the fog. However, the potentially vast amount of geographically distributed fog nodes may threaten the feasibility of the orchestration. On the other hand, fog nodes can exhibit highly dynamic workloads which may result in the orchestrator redistributing the services among them. Thus, there is also a need to dynamically support the network connections to those services independently of their location. Software Defined Networking (SDN) can be integrated within the orchestrator to carry out a seamless service management. To tackle both aforementioned issues, the S-HIDRA architecture is proposed. It integrates SDN support within a blockchain-based orchestrator of container-based services for fog environments, in order to provide low network latency and high service availability. Also, a domain-based architecture is outlined \marev{as potential scenario} to address the geographic distributed nature of fog environments. Results obtained from a proof-of-concept implementation assess the required functionality for S-HIDRA.
\end{abstract}

\begin{keyword}
Distributed systems \sep Fog computing \sep Software defined networking \sep Blockchain \sep Smart contracts \sep Resource orchestration



\end{keyword}

\end{frontmatter}

\section{Introduction}
\label{sec:intro}
Fog computing is an evolving paradigm that combines cloud computing and Internet of Things (IoT)~\citep{velasquez2018orchestration}, and places resources (computing power, storage, and memory capacity) closer to the edges of the network. In particular, fog computing relies on fog nodes with limited resources to create distributed cloud services on widely distributed edge networks. The time has come to consider decentralization and distribution for the future generation computing of IoT services and, more importantly, a distributed architecture for the orchestration of container-based \marev{services} running in fog computing to ensure resilience and fault-tolerance~\citep{ccek82022}. In this respect, blockchain provides a natural solution for fog computing decentralization, as well as to overcome some fog orchestration challenges such as security and auditability~\citep{cech2019blockchain}.

The blockchain technology has overcome its initial application in cryptocurrencies to reach over other application domains, which harness its immutable, transparent and available information storage capability. Moreover, smart contracts over blockchain can be seen as a distributed computer that is programmed to achieve different goals depending on the application. This type of digital contracts cannot be modified because their behavior itself (i.e. a set of promises) is stored as an irreversible transaction in the blockchain. Smart contracts are executed independently by each node participating in the blockchain. This fact is harnessed in our previous proposal, HIDRA~\citep{hidra2021}, to implement a truly distributed orchestrator for fog nodes. More precisely, \marev{this work is aimed at the orchestration of containers running in local fog environments} leveraging the adoption of both blockchain networks and lightweight virtualization technologies.

To address architectural changes in the fog virtualization infrastructure and the issues associated with the network communication, \marev{we propose S-HIDRA, an architecture to orchestrate services in broader Fog-IoT environments. Thus, S-HIDRA implements a novel network protocol to dynamically optimize the configuration of network devices (switches and routers) in real time, in order to improve the orchestration of containerized services. In this paper, we focus on the network efficiency of a decentralized fog architecture for Fog-IoT environments leveraging the adoption of Software Defined Networking (SDN)}. SDN is a technology that provides greater programming and flexibility for networks by the separation of the control plane from the data plane~\citep{priyadarsini2021}. \marev{By adding SDN capabilities, S-HIDRA streamlines our previous work to avoid static management of network traffic among fog nodes and IoT devices, thereby dynamizing network traffic towards the orchestrated containers.} This enables a programmable network architecture and allows traffic to be managed according to the network state (e.g. allowing load balancing, replication, and migration of containerized \marev{services}, etc). Also, in order to address the geographical distribution of fog environments, we outline a global fog architecture that harnesses S-HIDRA to segment scenarios with a vast amount of end-devices and fog nodes distributed along several geographical locations into different domains, coordinated by means of a global blockchain.

In summary, our key contributions in this paper are as follows:
\begin{itemize}
    \item \marev{We propose S-HIDRA, a novel SDN-based fog computing architecture that leverages the cooperation of the network layer with the container orchestration techniques to improve both the Quality of Service (QoS) and availability of the services supported by the system. S-HIDRA ensures data integrity and availability thanks to the blockchain-based orchestration.}
    \item \marev{We suggest the use of S-HIDRA in open and geographically broader fog computing scenarios that segment devices by domains such as smart buildings or campuses}.
    \item \marev{We design and implement an intra-domain protocol to orchestrate container services in distributed fog architectures using smart contracts and SDN virtual services.}
    \item \marev{We conduct the evaluation of S-HIDRA with a real testbed and real applications, aiming to bring realistic results.}
\end{itemize}

The rest of the paper is organized as follows. 
Section~\ref{sec:background} provides a brief background on the technologies involved in our proposal, and Section~\ref{sec:state_of_the_art} presents the related work. Then, Section~\ref{sec:shidra} introduces S-HIDRA as a distributed fog computing management architecture based on blockchain and SDN, and describes the components and operation model of the proposed architecture. In Section~\ref{sec:evaluation}, both the evaluation scenario and results are presented and discussed. Finally, conclusions are included in Section~\ref{sec:conclusions}.

\section{Background and building blocks}
\label{sec:background}
This section includes a brief outline of the context where this work is focused on, as well as the technologies used as building blocks in S-HIDRA. 

\subsection{Fog computing}
\label{sec:fog}
\marev{The term fog computing~\citep{bonomi2012fog} was coined as a contrast to cloud computing, in order to refer to the necessity of providing processing and storage in locations closer to data origins, mostly related to the IoT concept. Fog computing, and the likes such as edge or mist computing~\citep{mahmud2020fog}, allows for bandwidth savings and reducing latencies to/from the cloud, and can be used as an extension of the cloud for a wide range of latency-sensitive applications.}

A combination of different architectures, operating systems and resource specifications are very usual in current fog and IoT environments~\citep{habibi2020fog,Laroui2021,bilal2018}, including Single-Board Computers (SBC) which offer a great cost/performance ratio. In any case, there are usually limited storage and computing capabilities, especially when compared to the virtually infinite resource availability of cloud platforms.

Also, different applications with different requirements will coexist in the fog layer. Each of these applications should be isolated from the others for security reasons but also in order to provide them with their required Quality of Service (QoS). Light virtualization technologies such as containers can be harnessed to this end~\citep{morabito2017containers}. As a result, orchestration to track and manage the life cycle of containers becomes a key element within the fog layer.

\subsection{Blockchain}
\label{sec:blockchain}
Blockchain technology was first introduced in 2008 as the basis for Bitcoin~\citep{nakamoto2008bitcoin}. Its initial application in cryptocurrencies was overcome after the introduction of the Ethereum platform in 2013~\citep{buterin2013ethereum}, which included support to smart contracts~\citep{szabo1996smartcontracts}. These enabled the application of blockchain to a wide range of fields because arbitrary business logic can now be stored in the blockchain and autonomously executed when certain conditions are met~\citep{aggarwal2019}. Blockchain technology can be described as an append-only, sequential, immutable distributed ledger, where data are stored as transactions. Copies of this ledger are distributed among the participating nodes. They must agree on which data are added to the blockchain and in which order, by means of consensus algorithms~\cite{Kaur2021}.

Ethereum implements its own underlying peer-to-peer (P2P) network which connects all participating nodes on the blockchain. The blockchain transactions (or state transitions) are processed through the Ethereum Virtual Machine (EVM), which is able to execute the smart contracts. These are developed with high-level Turing-complete languages such as Solidity. Once the smart contracts are deployed, they can execute on every node of the blockchain and modify its global state according to their internal logic. This mimics the behavior of a global distributed computer, where every node in the blockchain participates independently and equally in updating the global system state, i.e. the single version of the truth.

\subsection{Software Defined Networking (SDN)}
\label{sec:sdn}
SDN is an approach to network management based on the separation of the data plane from the control plane~\citep{saraswat2019,jammal2014}. The Open Networking Foundation defines for SDN three planes for management of a network infrastructure. First, the \emph{data plane} composed of network devices (routers, switches, etc), whose function is the forwarding of frames and messages. All these devices lack intelligence and base their actions on the contents of their flow tables. The \emph{control plane} is formed by a centralized controller that interacts with the application plane in order to make decisions on data traffic and update the flow tables of the interconnection devices accordingly. Finally, the \emph{application plane} is composed of heterogeneous software responsible for implementing functions that have traditionally been executed in the network devices, such as routing algorithms, firewalls, load balancers, etc. This software makes the decisions that result in the entries in the flow tables. 

Each of these planes communicates with its adjacent through an interface. Thus, the application plane communicates with the control plane through the \emph{northbound interface}. Although this is not standardized, the most common controllers, such as ONOS~\citep{onos} or OpendayLight~\citep{opendaylight}, have their own RESTful APIs for integration. The communication between the control plane and the data plane uses the \emph{southbound interface}, which is standardized, with several standards available, such as OpenFlow~\citep{openflow}, NETCONF~\citep{netconf} or OVSDB~\citep{ovsdb}. Most controllers can use all of them for interaction with the data plane.

\section{State of the art}
\label{sec:state_of_the_art}
In this section, relevant contributions related to the topics addressed in this work are reviewed. More precisely, the focus is placed on recent proposals that combine fog computing, blockchain and/or SDN.

Regarding the integration of blockchain in fog/edge architectures, many proposals aim at enhancing the security of decentralized environments for IoT applications as well as efficiently handling distributed IoT data to provide integrity, authentication and privacy~\citep{fan2021sbbs,ali2018applications,tuli2019fogbus,baniata2020survey,Yang2019,Singh2020}. Smart contracts can also be deployed between clients and service providers to validate Service Level Agreements (SLA) compliance for on-demand resource usage, as in~\cite{Kochovski2020a}. As an example, in BlockEdge~\citep{Kumar2020}, the authors propose to use blockchain as a means for auditing the different processes involved in an industrial IoT application.

Focusing more specifically on resource management, the EdgeChain framework~\citep{pan2019edgechain} proposes a credit-based resource management system based on a permissioned blockchain and a currency system. The behavior of the IoT devices and rule enforcement is regulated by means of smart contracts. The blockchain is also used for secure data logging and auditing.

There are also some proposals that include both blockchain and SDN in fog/edge architectures. The blockchain-based distributed cloud architecture presented in~\cite{Sharma2018} includes controller fog nodes that enable SDN at the edge of the network to provide IoT services. The evaluation is performed by simulation and is focused on the response time of end clients compared to the cloud model.

The MultiChain blockchain framework is integrated into a low-cost fog computing environment in~\cite{cech2019blockchain}. Their fog layer includes controller nodes and gateways with SDN capabilities, integrated into a hierarchical architecture. However, the blockchain is used just as a mean to store and share confidential data from sensors, rather than as a cornerstone piece in the scheduling process together with SDN.

Other works focus on researching the security, existing attacks and mitigation mechanisms related to SDN technology~\citep{cui2021}. The authors in~\cite{AbouElHouda2019} tackle the problem of Distributed Denial of Service (DDoS) attack mitigation in SDN with a two-level strategy. DDoS patterns are identified within a domain (i.e. autonomous system) by applying machine learning techniques, while blockchain is harnessed for coordination between domains. To this end, Ethereum smart contracts are deployed, enabling multiple SDN-based domains to securely collaborate and transfer attack information in a decentralized manner. The set of devised smart contracts was tested over a personal test blockchain (Ganache) and also deployed over the Ropsten Ethereum testnet, although the authors only provide a cost analysis of the gas consumed by the contracts. Alternatively, the main focus of their evaluation is the capability to detect a DDoS attack within a domain. Another prominent research paper about the SDN security field employing blockchain technology is~\cite{rathore2019}. This work proposes a security architecture and detection mechanism for DDoS, ICMP flooding, and TCP flooding attacks based on a deep learning algorithm. The authors also implement the proposed security architecture using the Ethereum blockchain and the Mininet emulator in order to validate the attack detection scheme.

In~\cite{Misra2020} the authors propose a distributed SDN architecture for a fog-enabled, resource-constrained IoT network, where a private blockchain is used as a means for the coordination among different SDN controllers. Flow rules are stored on the blockchain so that they cannot be tampered. They are also available to any fog node, which might become eventually a SDN controller. The proposal is deployed on a testbed based on Raspberry Pi, which acts as fog nodes and SDN controllers. Also, systems with $i5$ processors are used as miners, which adds complexity to the proposal.

Several SDN controllers deployed in the cloud and in the fog are connected through a private blockchain in~\cite{GuhaRoy2021} in order to detect cyber attacks applying deep learning techniques. More precisely, the blockchain is harnessed to enhance failure tolerance among cooperating SDN controllers and provide a decentralized attack identification scheme. A similar approach is applied in~\cite{Luo2020}, where the blockchain is used again to synchronize the state among different SDN controllers. In this case, the focus of the work is the consensus algorithms and their implementation efficiency (in terms of energy consumption on the fog nodes) using deep reinforcement learning. This proposal is evaluated by means of simulation of the consensus protocols.

Thus, according to the reviewed literature, very few authors have worked on the resource management of fog/edge architectures by integrating blockchain and SDN technologies. As mentioned above, the main reasons for the integration are dealing with security issues and coordination among different SDN controllers. In this paper, we propose the S-HIDRA architecture as a blockchain-based, operating system-level virtualized resource management system with integrated network management based on SDN. That is, the main orchestration mechanism is based on blockchain smart contracts, which commands the SDN controller to perform the necessary actions for network connectivity.

\section{Proposed architecture: S-HIDRA}
\label{sec:shidra}
\marev{The main goal of the proposed architecture is to abstract the orchestration of container-based services and network services from administrators, application designers, and the fog nodes to provide QoS to connected IoT devices. Thus, S-HIDRA aims at combining the high degree of dynamism and programmability of SDN with secure, fault-tolerant, autonomous and auditable blockchain-based resource orchestration.}

To reach this goal, we propose distributed scenarios divided into different management and orchestration domains, \marev{such as smart buildings composed of unrelated companies or departments, smart campuses involving different schools or research groups, or distributed computing platforms. Domains in these scenarios} are interconnected at the network level and share a public peer-to-peer distributed ledger or blockchain to maintain a global state. Moreover, each domain internally maintains a local state to perform service orchestration tasks and to programmatically control network traffic to these services.

The following sections describe the proposed architecture in detail. First, Section~\ref{sec:overview} introduces the domain-based distributed architecture and the inter-domain and intra-domain characteristics. Section~\ref{sec:key_components} describes the components involved in a domain and how these components are related. Finally, Section~\ref{sec:workflow_protocol} details the orchestration workflow carried out by a domain that performs dynamic management of network traffic.

\begin{figure}[ht]
    \centering
    \includegraphics[width=8.75cm]{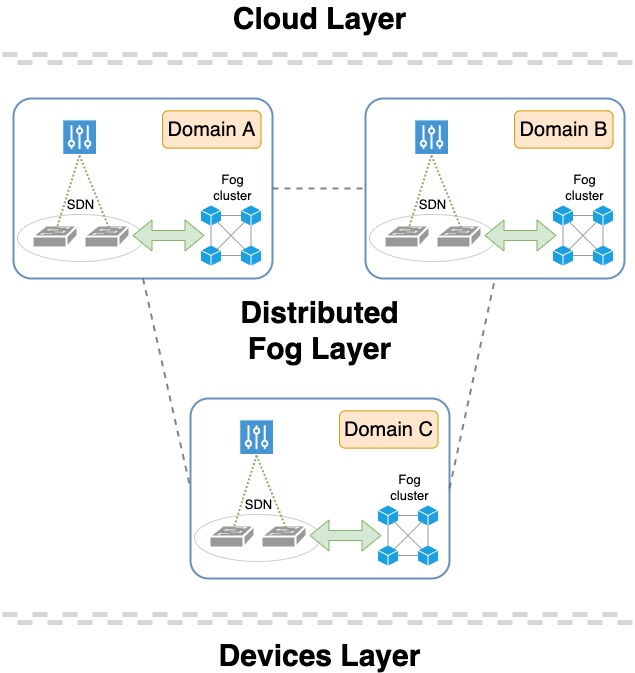}
    \caption{S-HIDRA global overview.}
    \label{fig:global_overview}
\end{figure}

\begin{figure*}[ht]
    \centering
    \includegraphics[width=12cm]{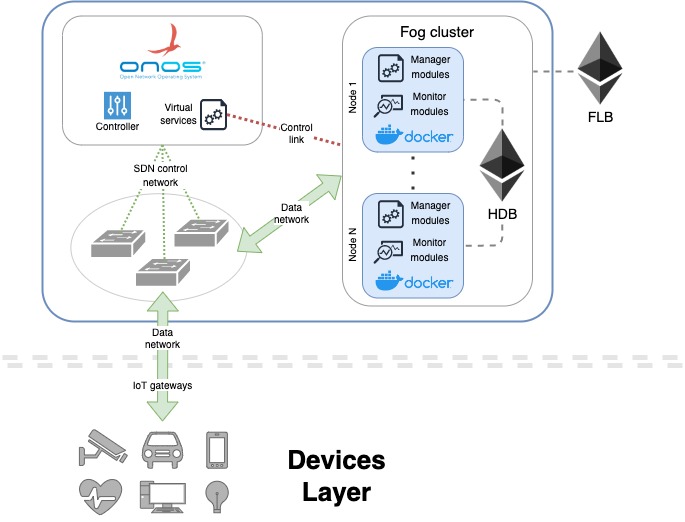}
    \caption{S-HIDRA domain overview.}
    \label{fig:domain_overview}
\end{figure*}

\subsection{Global overview}
\label{sec:overview}
\marev{Figure~\ref{fig:global_overview} depicts the high-level layered architecture of S-HIDRA. In the proposed architecture, the Distributed Fog Layer is composed of several domains, allowing computing and network operations to be distributed throughout the fog layer. Each domain consists of a local cluster of fog nodes where S-HIDRA dynamizes network traffic towards containerized services by adding SDN capabilities to the existing blockchain-based orchestration platform~\cite{hidra2021}.}
This enables a programmable network architecture and allows traffic to be managed according to the network state (e.g. allowing load balancing, replication, and migration of containerized \marev{services}, etc).

\begin{figure}[ht]
    \centering
    \includegraphics[width=9cm]{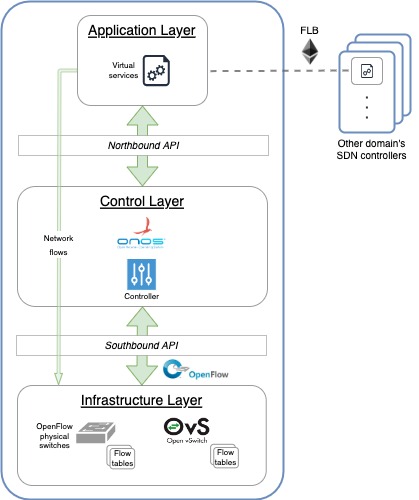}
    \caption{\marev{SDN plane view.}}
    \label{fig:sdn_plane}
\end{figure}

\begin{figure}[ht]
    \centering
    \includegraphics[width=8.7cm]{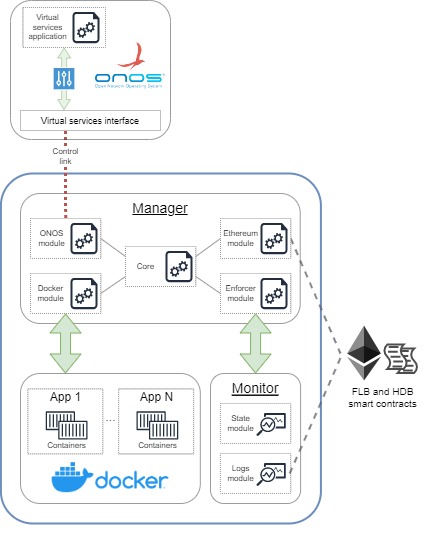}
    \caption{Orchestration plane view.}
    \label{fig:orchestration_plane}
\end{figure}

Regarding the inter-domain plane, domains are interconnected and share a global blockchain called Fog Layer Blockchain (FLB). The FLB could be a private permissioned blockchain deployed in a consortium manner by the system domains, or even an open permissionless blockchain accessible worldwide. In any case, the FLB deploys the Inter-Domain Registry (IDR) functional module and smart contract that manages the global domain state, i.e. the shared public metadata that domains require from other domains. Examples of metadata managed by the IDR could be domain registration data, information about inter-domain migrations, IP address and port of publicly exposed \marev{services}, domain reputation scores, node blacklists, etc.

\marev{In the proposed fog architecture, each SDN controller focuses on managing network devices in a single domain. For the distribution of network information between SDN controllers, we recreate the flat model or horizontal architecture described in~\cite{oktian2017sdn}. The inter-domain SDN behavior is governed at the application level by the FLB. Therefore, domain SDN controllers are able to interact with each other to record and query network information from containerized services and orchestration events that occur at intra-domain level.}

\marev{At intra-domain level, a domain is composed of (1) a cluster of fog nodes connected through a local permissioned blockchain called HIDRA Domain Blockchain (HDB), and (2) an SDN networking layer in which an SDN controller is responsible for managing a subset of Distributed Fog Layer network devices}. Regardless of the number of network devices and topology generated within a domain, an SDN controller can dynamize the traffic to/from fog nodes querying the local domain state stored in the HDB in a secure and decentralized manner. The local domain state refers to the information and metadata kept by \marev{the S-HIDRA's functional modules (more details in Section~\ref{sec:key_components}). Note that these metadata are isolated from other domains. Only registered devices can participate in a cluster}, being able to query and agree on the local domain state. This improves privacy and scalability within a domain by partitioning blockchain transactions and data stored by each HDB.

Figure~\ref{fig:domain_overview} represents an S-HIDRA domain overview containing the key components and required intra-domain connections. As introduced before, each domain is composed of an SDN controller and a cluster of fog nodes. Regarding the first, we propose the use of ONOS, a leading open source SDN controller designed for high availability, performance and scalability that allows the development and deployment of SDN applications in a modular way and provides features for distributed control. The SDN approach divides network traffic into control, data, and application planes. These planes can be seen in the figure. It is worth mentioning the control plane or \emph{SDN control network}, through which the SDN controller and domain switches interact, and the data plane or \emph{data network} indicating a bidirectional data traffic between the fog cluster and the Devices Layer through IoT gateways. Regarding the SDN application plane, we have developed an ONOS application that manages virtual services. In S-HIDRA, a virtual service represents, at SDN level, a containerized \marev{service} deployed and orchestrated in the system.

An S-HIDRA domain cluster is composed of $N$ fog nodes that perform tasks in a consensual manner to orchestrate resources and \marev{services} following the workflow protocol described in detail in Section~\ref{sec:workflow_protocol}. Both FLB and HDB blockchains are implemented using the Ethereum technology, so the smart contracts related to the proposed functional modules are executed through the EVM. Although the SDN approach divides traffic into different planes or networks, fog nodes belonging to the data plane have a \emph{control link} that allows them to send management instructions to the ONOS application plane, specifically to the virtual services application. This allows network traffic to be modeled according to the orchestration decisions made. Note that all system fog nodes implement the same software client required for inter-domain and intra-domain communication (i.e. communication with FLB and HDB blockchains, SDN controllers and Docker daemons), and for resource monitoring and rule enforcement.

\subsection{Key components}
\label{sec:key_components}
In order to increase the understanding of the proposed architecture and its structure, this section defines the S-HIDRA key components and their connections from two different points of view: \marev{the SDN plane view (detailed in Figure~\ref{fig:sdn_plane}) and the orchestration plane view (Figure~\ref{fig:orchestration_plane})}. 

Figure~\ref{fig:sdn_plane} shows the SDN architecture divided into the infrastructure, control and application layers and adapted to S-HIDRA requirements. The infrastructure layer refers to network devices, whether physical or virtual switches, that reactively store flow rules based on the state desired by S-HIDRA. Then, the control layer is responsible for performing the SDN logic and managing the network topology and network devices through southbound APIs such as OpenFlow. Finally, the control layer interacts via northbound APIs with the application layer. This layer is composed of applications capable of dynamically and programmatically managing the network using the resources of SDN controllers deployed on the control layer. Specifically, S-HIDRA implements an SDN application that manages virtual services as a network abstraction for containerized \marev{services}. The virtual services application is the entry point to the management of SDN network flows by fog nodes.

\marev{The orchestration plane view} is shown in Figure~\ref{fig:orchestration_plane}. This view exposes the key components of S-HIDRA nodes and the connection between these nodes and the SDN plane. \marev{In the figure, we also distinguish some components related to our previous research work HIDRA~\citep{hidra2021}. HIDRA establishes the basis of the domain-based distributed architecture proposed by S-HIDRA. A HIDRA cluster sets a local P2P network to share and synchronize the cluster state. To this end, a permissioned Ethereum blockchain is deployed over the P2P network, enabling the execution of smart contracts. On the other hand, HIDRA monitors and isolates node resources (i.e. CPU, memory, disk storage, etc) through the Monitor component and Docker virtualization service, respectively. Other key components, such as the Manager, had limited functionality for orchestrating resources in HIDRA.}

\marev{Focusing on S-HIDRA, we detail below the key components required in the distributed SDN-based orchestration:}
\begin{itemize}
    \item \textbf{SDN controller}. In S-HIDRA, each domain deploys an ONOS controller that manages the domain network devices. \marev{At inter-domain level, synchronization of network information is achieved via the FLB at the SDN application level. Thus, ONOS controllers in S-HIDRA are interconnected forming a distributed SDN data store. On the other hand, at intra-domain level, it is possible to form local clusters by replicating instances of ONOS controllers and leaving one of these instances as the master SDN controller and the others taking a standby role~\citep{onos2016cluster}. This avoids single points of failure and increases the availability of SDN functions within S-HIDRA domains.}
    \item \textbf{Virtual services application}. The ONOS platform supports the development of applications independently of its core system. ONOS applications extend SDN capabilities and allow greater control over network traffic. In S-HIDRA, most of the network traffic refers to Devices Layer requests and containerized \marev{services} responses. Since the orchestrator mechanism migrates \marev{services}/containers according to certain enforcement mechanisms (e.g. nodes resource usage, nodes geolocation, nodes reputation, etc), dynamic traffic management is useful to ensure communications between devices and fog nodes. Thus, each S-HIDRA domain deploys a local instance of the virtual services application. This ONOS application allows to abstract the network configuration from the deployed containerized \marev{services}. A virtual service is linked to a single containerized \marev{service} and it is responsible for allocating and configuring an IP, port, and protocol (TCP/UDP) in order to expose the containerized \marev{service} within a domain, regardless of the fog nodes this \marev{service} is running on. The virtual services application also provides other features such as replication and load balancing between different replicas of the same containerized \marev{service} (which may be running on different fog nodes).
    \item \textbf{Virtual services interface}. Fog nodes make distributed decisions about where to deploy the containerized \marev{services}. At the same time, fog nodes register and manage virtual services within an S-HIDRA domain. \marev{Therefore, a link interface is required to enable S-HIDRA nodes to send and query network information to satisfy inter-domain and intra-domain SDN states.} The ONOS platform allows the implementation and exposure of custom REST APIs that intermediate between ONOS applications and the outside world. Note that for security reasons, the \emph{control link} between fog nodes and the virtual services interface is isolated from the \emph{SDN control network} (more details about this in Section~\ref{sec:evaluation}).
    \marev{\item \textbf{Functional modules/smart contracts.} S-HIDRA divides the control logic into five functional modules, each of them implementing one or more EVM smart contracts that distribute the system logic among all participants. Thus, fog nodes employ blockchain transactions to share and query data and exchange control messages related to orchestration tasks. These functional modules are: (1) the \emph{Distributed Device Registry (DDR)} which registers and authenticates fog nodes, (2) the \emph{Distributed Event Logger (DEL)} responsible for managing orchestration events, (3) the \emph{Distributed Container Registry (DCR)} focused on storing the historical and desired state in terms of resources and containerized \marev{services}, (4) the \emph{Distributed Reputation System (DRS)} which calculates the reputation score of nodes according to their actions, and (5) the \emph{Inter-Domain Registry (IDR)} to globally share data generated by the intra-domain workflows. Modules (1)-(4) are executed in isolation through the HDB, and module (5) is shared among domains via the FLB.}
    \item \textbf{ONOS module}. \marev{The software client executed by each S-HIDRA node requires a module to interact with the SDN capabilities. The ONOS module located inside the Manager component is responsible for managing requests} to the virtual services application through the virtual services interface. During the process of deploying, migrating or deleting containerized \marev{services} within a domain, the workflow protocol described in Section~\ref{sec:workflow_protocol} determines when and which fog node requests the ONOS module to create, modify or delete virtual services linked to a containerized \marev{service}.
    \item \textbf{Docker module}. Note that deployment/migration and execution times of a container depend on its type and complexity. In S-HIDRA, these times are critical for the synchronization of containers and virtual services: during a container migration process, the virtual service linked to the containerized \marev{service} is the same, so there may be small downtimes until the migrated container is active and healthy on the destination node (i.e. on the solver node). This would affect the availability of containerized \marev{services}. Therefore, the Docker module includes a \emph{container health check system} that monitors the internal state of containers and notifies the state to fog nodes. This feature is also reflected in the workflow protocol.
\end{itemize}

\begin{figure*}[ht]
    \centering
    \includegraphics[width=16.7cm]{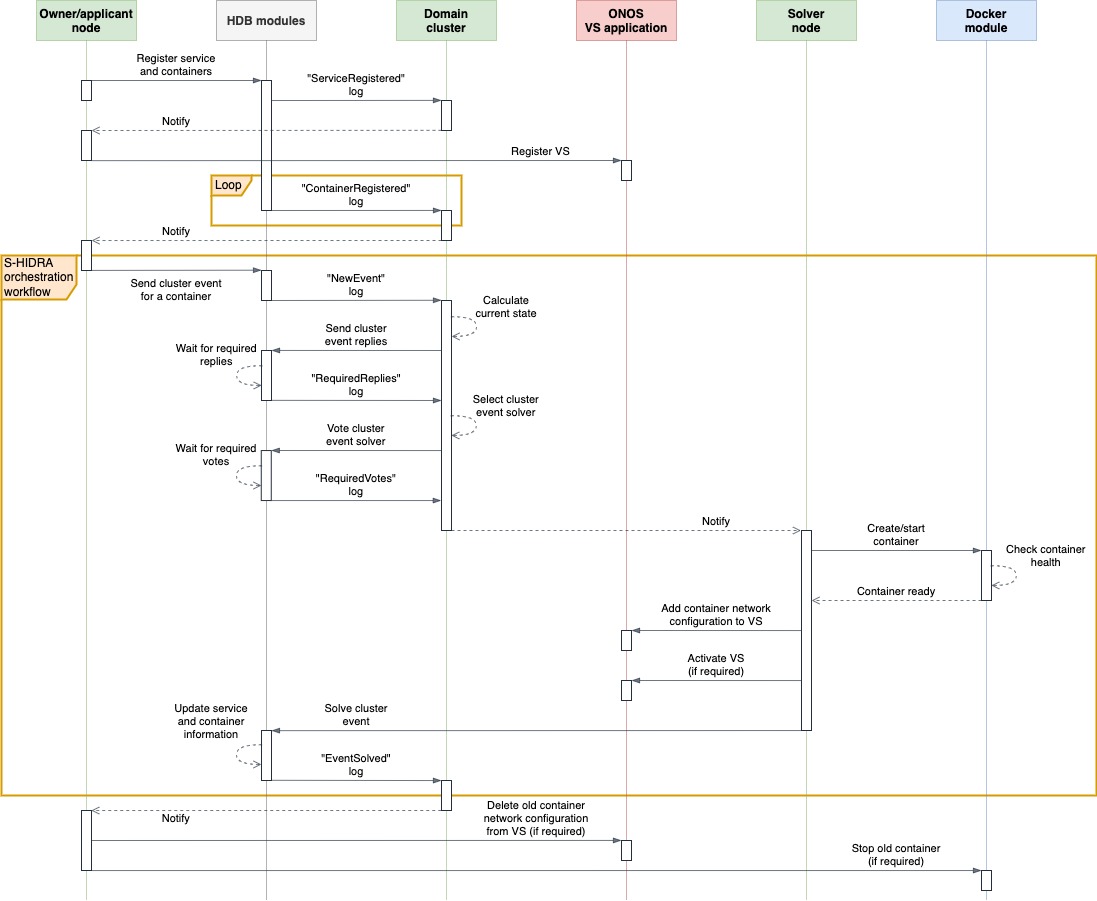}
    \caption{\marev{S-HIDRA workflow protocol.}}
    \label{fig:workflow_protocol}
\end{figure*}

The Ethereum and logs modules, shown in Figure~\ref{fig:orchestration_plane}, are the only modules that directly interact with the blockchain. The Ethereum module is responsible for signing and sending transactions and interacting with the S-HIDRA functional modules. The logs module listens for any EVM asynchronous log emitted by a smart contract in response to inbound transactions. S-HIDRA requires the integration of these modules with both the HDB and the FLB to connect fog nodes at intra-and at inter-domain levels, respectively. Regardless of this requirement, the following sections focus on describing and evaluating the functionality and performance of an S-HIDRA domain, delving into the orchestration workflow protocol and the connections between the S-HIDRA components. The evaluation of the proposal at inter-domain level depends on the use case, the type of FLB, and the network topology, issues that are out of the scope of this paper.

\subsection{Workflow protocol details}
\label{sec:workflow_protocol}
In this section, we delve into the S-HIDRA intra-domain workflow protocol and detail the container orchestration and distributed decision-making processes. Figure~\ref{fig:workflow_protocol} shows the steps required to orchestrate containerized \marev{services} at intra-domain level. The entities participating in the workflow protocol are shown at the top of the figure. From left to right, (1) a domain fog node acting as a containerized \marev{service} owner or as an applicant node with problems, (2) the HDB functional modules/smart contracts, (3) a cluster representing all fog nodes of a domain, (4) the SDN capabilities including the ONOS virtual services application and virtual services interface, (5) the solver node selected by the cluster to carry out the tasks required by the owner/applicant node, and finally, (6) the Docker module.

S-HIDRA domains are capable of synchronizing information through cluster events using the DEL functional module and the log exchange protocol implemented in the HDB smart contracts. Although the DEL module allows many types of cluster events, from now onwards we focus on the container deployment and migration processes. 
The deployment process starts with a fog node that requires the creation of a new containerized \marev{service}. In this case, the node assumes the owner role and takes control of this \marev{service}, regardless of which cluster node executes the \marev{service's} containers. To link the containerized \marev{service} with an SDN virtual service, first, the owner node registers the metadata of the new \marev{service} and its containers in the DCR module. Once the metadata registration transaction has been sent and validated by the HDB, multiple asynchronous logs are generated: an \marev{\emph{ServiceRegistered}} log that notifies the owner in order to register a new virtual service in ONOS linked to the containerized \marev{service}, and a \emph{ContainerRegistered} log for each \marev{service} container.

\begin{figure*}[ht]
    \centering
    \includegraphics[width=11cm]{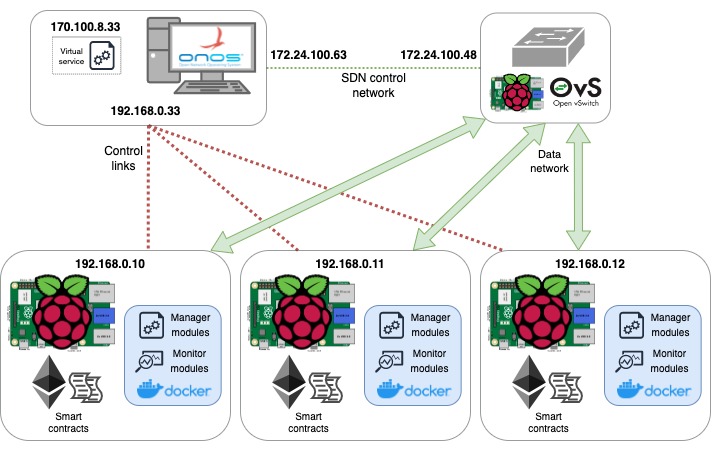}
    \caption{S-HIDRA testbed.}
    \label{fig:shidra_testbed}
\end{figure*}

Thus far, the containerized \marev{service} has been registered but none of its containers have been orchestrated or executed yet. Besides this, the registered virtual service remains inactive. The next step is to execute an S-HIDRA orchestration workflow for each registered container. Note that a container migration process would also start at this point. In this case, the node assumes the applicant role after detecting a problem with its resources or running containers. In both processes, an S-HIDRA orchestration workflow begins with a node sending a DEL event related to a container (i.e. the cluster event in Figure~\ref{fig:workflow_protocol}). Once the event transaction has been sent and validated, the DEL smart contract throws a \emph{NewEvent} log that notifies the other cluster nodes. Each node then generates a report containing its current state and sends it as a reply to the DEL event. Until the required event replies are collected, the DEL contract does not throw a \emph{RequiredReplies} log that allows the orchestration workflow to continue. The number of replies required depends on the initial cluster configuration.

Now that the cluster has been synchronized and all fog nodes know the current state, it is possible to elect the solver node that will host the container in a decentralized manner. To this end, each cluster node selects a solver node considering the current cluster state and sends a vote transaction to the DEL contract. Similar to the wait-for-replies phase, the orchestration workflow waits until the DEL module has received the number of votes required by the initial cluster configuration. Once the votes have been collected, the DEL contract notifies the elected solver node via a \emph{RequiredVotes} log.

At this point, the cluster has the information required to create the owner node container or replicate the applicant node container. In any case, the solver node instantiates the container using the Docker module and configures it according to the metadata registered by the owner node. In a container migration process, the cluster will be executing two instances of the applicant node container for a short time \marev{(redirecting for now all device requests through the old instance)}. Regardless of the event type, the Docker module's health check system is responsible for monitoring the health of the container. When the system reports a healthy container state, the solver node sends the network configuration of the new container instance (node IP, exposed port, and protocol) to the virtual service previously registered in ONOS. \marev{Note that not all containers need to register their network configuration in a virtual service, some containers work internally linked via Docker. Subsequently, once the virtual service has been updated, the solver node activates it if required.}

To complete the cluster event, the solver node sends a solve transaction to the DEL contract. This transaction can only be sent by the solver node. A solve transaction closes a cluster event but also changes the cluster state depending on the event sent. \marev{Although domain devices could perform requests to the containerized service just after activating the virtual service (before solving the event),} the HDB functional modules are the only ones responsible for maintaining the desired cluster state. It means, for example, that the DCR decides which containerized \marev{services} are active, inactive, or deleted, regardless of the state of the ONOS virtual services application.

To finalize the S-HIDRA orchestration workflow, the DEL contract notifies the cluster nodes via an \emph{EventSolved} log that the cluster event has been solved. At this point, the cluster nodes may carry out multiple actions in order to seal the new cluster state: \marev{nodes or service availability checks, send metadata to the FLB and IDR module, reputation calculations, etc. The applicant node may also perform cleaning actions related to the problematic/old container instance.} Thus, the virtual service now only redirects network traffic from devices to the new container hosted by the solver node, allowing the applicant node to release resources.

\section{Proposal evaluation}
\label{sec:evaluation}
This section evaluates the functionality and other non-functional attributes of the S-HIDRA workflow protocol described in the previous section. To reach this goal, we have implemented a testbed that emulates an S-HIDRA \marev{real-world} domain. Figure~\ref{fig:shidra_testbed} shows the key components of a domain composed of three fog nodes interconnected through an OpenFlow switch controlled by an ONOS controller 2.6.0. Both the OpenFlow switch and the fog nodes are implemented in SBC devices with different specifications. In particular, the OpenFlow switch is based on a Raspberry Pi 3 Model B with 1.2GHz quad-core CPU and 1GB of memory, while the cluster nodes are Raspberry Pi 4 with 1.5GHz quad-core CPU and 4GB of memory. Using SBC devices to build the testbed allows us to evaluate the S-HIDRA workflow protocol on devices with limited resources, demonstrating the proposal feasibility even in IoT constrained environments. The Raspberry Pi acting as the switch executes Open vSwitch (OVS) 2.10.7 in order to virtualize the networking capabilities of an SDN switch.

\begin{figure}[ht]
    \centering
    \includegraphics[width=9cm]{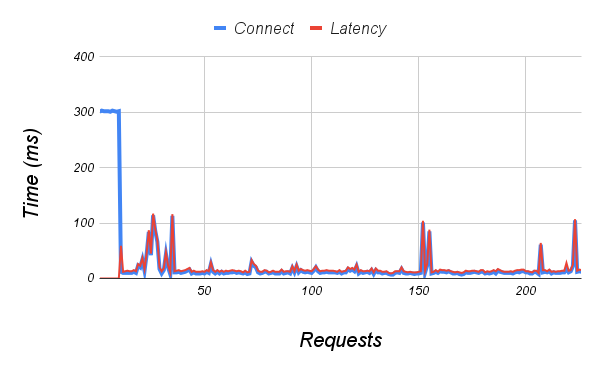}
    \caption{Network measurement during an NGINX container deployment.}
    \label{fig:nginx_deploy}
\end{figure}

\begin{figure}[ht]
    \centering
    \includegraphics[width=8.75cm]{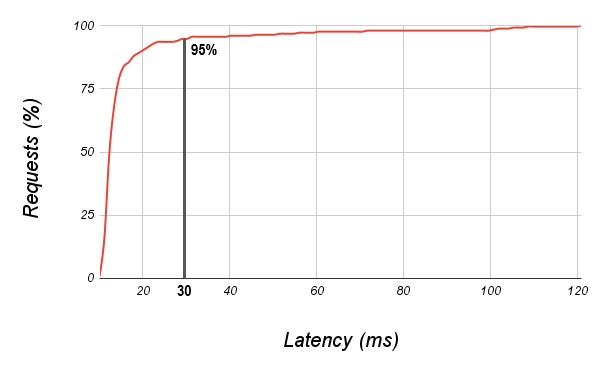}
    \caption{Percentage of requests and latency during an NGINX container deployment.}
    \label{fig:nginx_latency_per}
\end{figure}

\marev{As detailed in previous sections, each node executes the HDB (based on Geth 1.10.11), the light virtualization service (using Docker 20.10.10), and the Manager and Monitor components. The HDB smart contracts are written in the Solidity programming language, while the Manager and Monitor components are written in Go for a smooth integration with Geth and Docker.}

\begin{figure*}[ht]
    \centering
    \begin{subfigure}{8cm}
        \centering
        \includegraphics[width=1\textwidth]{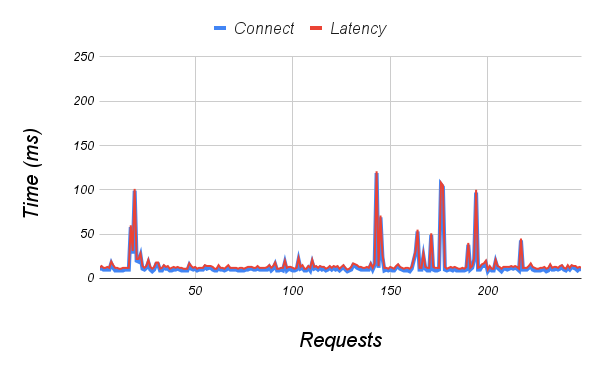}
        \caption{NGINX container.}
        \label{fig:nginx_migration}
    \end{subfigure}
    \begin{subfigure}{8cm}
        \centering
        \includegraphics[width=1\textwidth]{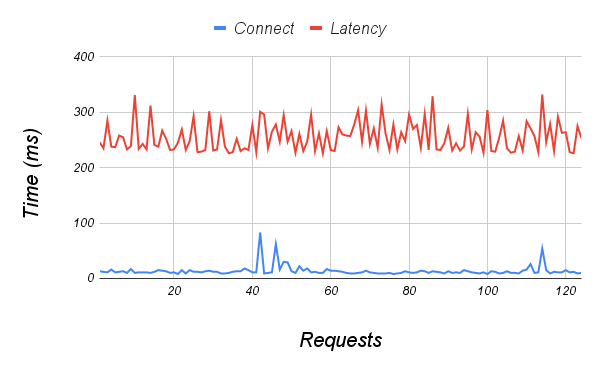}
        \caption{Nextcloud container.}
        \label{fig:nextcloud_migration}
    \end{subfigure}
    \begin{subfigure}{8cm}
        \centering
        \includegraphics[width=1\textwidth]{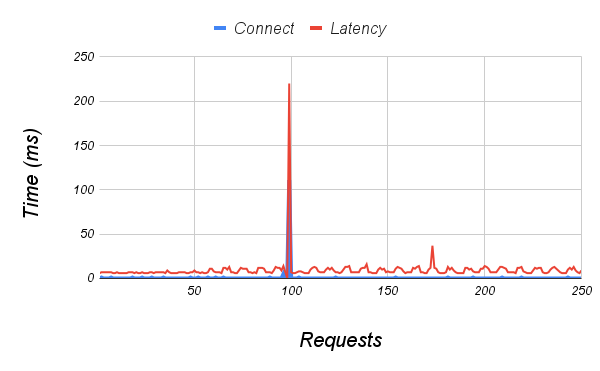}
        \caption{PostgreSQL container.}
        \label{fig:postgresql_migration}
    \end{subfigure}
    \caption{Network measurement during a container migration.}
    \label{fig:container_migrations}
\end{figure*}

The testbed deploys the SDN components described in Section~\ref{sec:key_components} on a Debian 10 virtual machine with 2.5GHz quad-core CPU and 4GB of memory. These components are the ONOS controller, the virtual services application and the virtual services interface, all written in the Java programming language. The network configuration of the SDN components and cluster nodes depends on the initial cluster configuration. We differentiate three different IP networks in the testbed: (1) the 192.168.0.0/24 network corresponding to the \emph{data network} that interconnects the cluster nodes and other devices from the Devices Layer, (2) the 172.24.100.0/23 network, pre-configured for the ONOS controller and the OVS, composing the \emph{SDN control network}, and (3) the custom network addresses representing the virtual services, e.g. 170.100.8.33. The initial cluster configuration also specifies the ONOS controller network address (192.168.0.33) that acts as management endpoint for the \emph{control links} between cluster nodes and the virtual services interface.

Once the testbed has been described, we proceed to discuss the tests carried out to evaluate the S-HIDRA workflow protocol and the results obtained. \marev{We want to evaluate the performance of the system for real use cases which are latency sensitive, like event processing IoT applications (i.e. healthcare, industry 4.0, etc). Moreover, we will focus on the availability of the containerized \marev{services}. Note that availability is an important metric in fog architectures especially when migration or re-scheduling of containers is performed.} Specifically, the tests verify the operation of an S-HIDRA domain during the container deployment and migration processes, and measure attributes such as network latency or \marev{service} availability. Note that the tests are performed from an observer cluster node and the required virtual services are registered in ONOS in advance for each test.

Figure~\ref{fig:nginx_deploy} shows the \emph{connect times} and \emph{latency times} during a container deployment process. \emph{Connect times} represent the time taken to establish the TCP connection with the endpoint, while \emph{latency times} include the \emph{connect times} plus the endpoint computation time and the time until after receiving the first response. In this first test, the deployed container is an NGINX web server. Each request sent to the container endpoint represents a new TCP connection, since the source port used by the observer cluster node is different. This means that for each request sent, a new SDN flow rule is generated and installed on the OVS. The results show \emph{connect times} peaks at the beginning of the test due to the 300ms timeouts that take place before the container exists. Once the container deployment instruction is sent and the S-HIDRA orchestration workflow is complete, the NGINX web server starts receiving requests successfully. In general, the \emph{latency times} are around 16ms, although there might be some peaks depending on the SDN components saturation. Figure~\ref{fig:nginx_latency_per} shows the percentage of requests and latency during the first test. The results obtained reveal that despite the peaks, a high percentage of requests will remain at low latency levels. Specifically, 95\% of requests will remain below 30ms.

The next tests are about the container migration process\marev{, as it is the most costly and critical orchestration task. It involves spinning up the service in the solver node and redirecting traffic flows to it. Moreover, the service cannot be stopped in the applicant node until it is up and running in the solver node. Thus, it becomes essential to monitor service availability while testing S-HIDRA.} In this case, we measure the \emph{connect times} and \emph{latency times} during the migration of different types of container: an NGINX web server, a Nextcloud file hosting server and a PostgreSQL database. This container variety shows that S-HIDRA works independently of the type of container orchestrated. Figure~\ref{fig:container_migrations} displays the requests sent in each container migration. The web server migration results are similar to the deployment process results, i.e. some peaks also occur that are independent of the orchestration workflow. In contrast, the Nextcloud and PostgreSQL container migration tests show different results. Regarding the Nextcloud file hosting server, the \emph{connect times} are around 13ms, but the \emph{latency times} increase up to 330ms. This is due to the internal processing that the container performs with each request received. 
On the other hand, query requests to the PostgreSQL container are sent over the same TCP stream, so during the test the \emph{connect times} are insignificant. Note that only two SDN flow rules are installed on the OVS in this test: one rule before and one rule after migration. This is reflected in Figure~\ref{fig:postgresql_migration}, where the \emph{latency times} remain around 8ms but there is a 220ms peak due to the TCP reconnection and the installation process of the new flow rule.

\begin{figure}[b!]
    \centering
    \includegraphics[width=9cm]{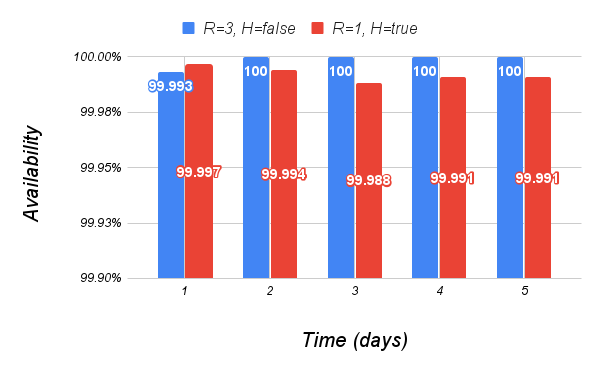}
    \caption{Availability per day according to request rate and container health check.}
    \label{fig:nginx_availability}
\end{figure}

\begin{table*}[t]
\centering
\caption{Availability test statistics.}
\label{table:availability_data}
\begin{tabular}{ccccccc}
\hline
 & \multicolumn{3}{c}{\textbf{R=3, H=false}} & \multicolumn{3}{c}{\textbf{R=1, H=true}} \\ \hline
\multirow{2}{*}{\textbf{Migrations}} & \textit{Per hour} & \textit{Per day} & \textit{Total} & \textit{Per hour} & \textit{Per day} & \textit{Total} \\
 & 66.06 & 1585.40 & 7927 & 46.43 & 1114.40 & 5572 \\ \hline
\multirow{2}{*}{\textbf{Requests}} & \textit{Per hour} & \textit{Per day} & \textit{Total} & \textit{Per hour} & \textit{Per day} & \textit{Total} \\
 & 1200 & 28800 & 144000 & 3600 & 86400 & 432000 \\ \hline
\textbf{Errors} & \multicolumn{3}{c}{2} & \multicolumn{3}{c}{34} \\ \hline
\textbf{Max. latency (ms)} & \multicolumn{3}{c}{1156} & \multicolumn{3}{c}{1103} \\ \hline
\textbf{Min. latency (ms)} & \multicolumn{3}{c}{8} & \multicolumn{3}{c}{7} \\ \hline
\textbf{Avg. latency (ms)} & \multicolumn{3}{c}{50.71} & \multicolumn{3}{c}{16.09} \\ \hline
\textbf{Std. Dev.} & \multicolumn{3}{c}{186.90} & \multicolumn{3}{c}{18.88} \\ \hline
\end{tabular}
\end{table*}

\begin{figure*}[t]
    \centering
    \begin{subfigure}{9cm}
        \centering
        \includegraphics[width=1\textwidth]{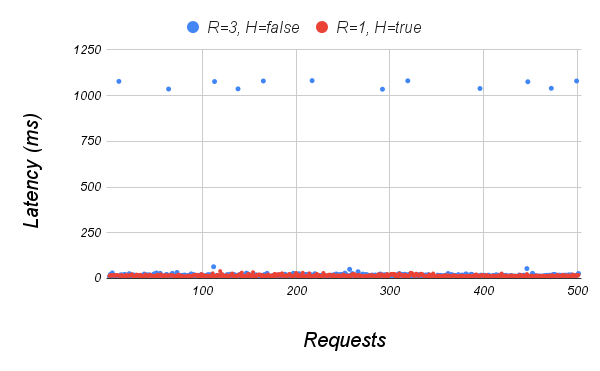}
        \caption{Short view.}
        \label{fig:nginx_latency_short}
    \end{subfigure}
    \begin{subfigure}{9cm}
        \centering
        \includegraphics[width=1\textwidth]{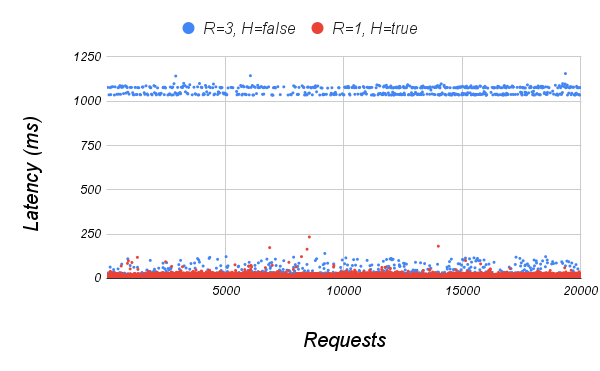}
        \caption{Long view.}
        \label{fig:nginx_latency_long}
    \end{subfigure}
    \caption{Requests latency according to request rate and container health check.}
    \label{fig:nginx_latency}
\end{figure*}

The remaining tests measure the availability of a containerized \marev{service} during successive container migrations over time. Figure~\ref{fig:nginx_availability} shows the percentages of successful requests sent to an NGINX web server container during two availability tests performed over five consecutive days. In these tests, different configurations were compared based on the time between consecutive requests sent to the container endpoint ($R$ parameter -- secs), and on the use of the health check system when the S-HIDRA orchestration workflow executes a container ($H$ parameter -- true/false). In both tests, the availability percentages are around 99.99\%. These results show that the orchestration workflow remains synchronized with the state of the ONOS virtual services application, minimizing request loss when orchestrating containerized \marev{services}.

Additionally, Table~\ref{table:availability_data} collects data from the tests performed, such as the number of migrations and requests sent, and some latency statistics. During the test configured with $R=3$ and $H=false$, 7,927 container migrations are performed while the observer cluster node sends 144,000 requests to the container endpoint. With $R=1$ and $H=true$, there are 5,572 migrations and 432,000 requests. The number of failed requests during the two tests was 2 and 34 respectively. Note that the number of failed requests depends on the $R$ parameter, but not on the $H$ parameter. The increase in failed requests in the second test is due to the increase in the sending rate (i.e. three times more requests sent). Despite the failed requests, in the second test the \emph{latency times} improved considerably, decreasing from an average latency of 50.71ms in the first test, to an average latency of 16.09ms.

Previous results show the feasibility of the S-HIDRA orchestration workflow, regardless of the number of requests sent to containerized \marev{services}. Figure~\ref{fig:nginx_latency} presents a detailed view of the \emph{latency times} during a slice of the availability tests. Both Figure~\ref{fig:nginx_latency_short} and Figure~\ref{fig:nginx_latency_long} show latency peaks around 1000ms with $H=false$. These peaks are due to the fact that in a container deployment/migration process, the workflow does not wait to receive a healthy container state, so the cluster permits sending requests to containers that are running but have not been initialized yet. The latency peaks depend on container type, so the higher the container complexity, the longer the initialization time and request latency, even increasing the number of failed requests. In contrast, with $H=true$, until the container reports a healthy state, the virtual services application and the OVS redirect requests to other container instances (i.e. to other container replicas or, in the worst case, to the applicant instance).

\section{Conclusions}
\label{sec:conclusions}
In this paper we have presented S-HIDRA, a domain-based architecture focused on orchestration of resources as containerized \marev{services} in decentralized fog computing environments. We based this proposal on a previous research work in which a first version of the blockchain-based HIDRA orchestrator was proposed. S-HIDRA has been designed to address fog computing environments that are geographically broader and segmented into device/node sets as domains. In addition, S-HIDRA inherits important features from blockchain technology such as immutability, availability or transparency.

In order to dynamize and programmatically manage network traffic to decentrally orchestrated containerized \marev{services}, we have proposed the use of SDN capabilities and their integration in the HIDRA orchestrator and S-HIDRA domains. We also implement a testbed that emulates an S-HIDRA domain to evaluate the feasibility of the proposal. The results obtained show a proper orchestration as well as low latency levels and high availability of containerized \marev{services} during the container deployment and migration processes.

In future work, we will study specific use cases where we could apply the S-HIDRA proposal, such as smart cities or campuses. These use cases will allow us to measure the feasibility of the proposal at inter-domain level. We also intend to design an advanced reputation system capable of quantifying the behavior of nodes and domains in order to make more complex decisions when orchestrating containerized \marev{services} in S-HIDRA. This reputation system will be based on the detection of \marev{spam, Sybil~\citep{douceur2002} and free-riding~\citep{locher2006} attacks to manage the reputation scores}. \mirev{Finally, different scheduling mechanisms for fog computing environments should be studied in depth. In this area, we plan to design and implement a geolocation-based mechanism to orchestrate containerized services using metrics such as the Haversine distance among fog nodes.}

\section*{Acknowledgments}
\marev{This work has been funded by MCIN/AEI/10.13039/501100011033 and by European Regional Development Fund (ERDF), ``A way to make Europe'' (ref. PID2021-123627OB-C52),} 
\mirev{and under GC-020-017 grant, funded by the Regional Government of Castilla-La Mancha for Consolidated Research Groups}.



\bibliography{bibliography}
\bibliographystyle{elsarticle-num-names}






\end{document}